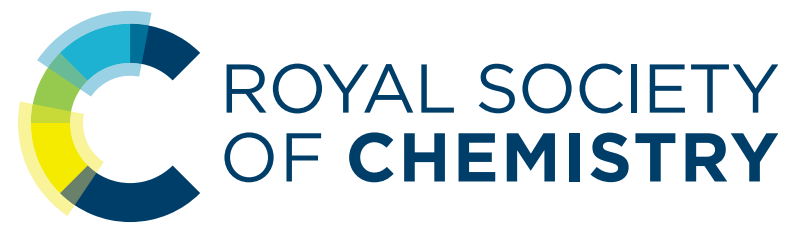

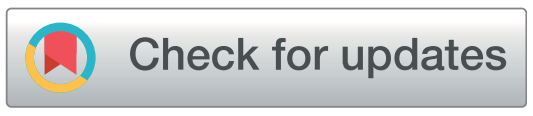

# Application-specific machine-learned interatomic potentials: exploring the trade-off between DFT convergence, MLIP expressivity, and computational cost

Ilgar Baghishov, [*ac] Jan Janssen, [ad] Graeme Henkelman [bc] and Danny Perez [*a]

Machine-learned interatomic potentials (MLIPs) are revolutionizing computational materials science and chemistry by offering an efficient alternative to *ab initio* molecular dynamics (MD) simulations. However, fitting high-quality MLIPs remains a challenging, time-consuming, and computationally intensive task where numerous trade-offs have to be considered, *e.g.*, How much and what kind of atomic configurations should be included in the training set? Which level of *ab initio* convergence should be used to generate the training set? Which loss function should be used for fitting the MLIP? Which machine learning architecture should be used to train the MLIP? The answers to these questions significantly impact both the computational cost of MLIP training and the accuracy and computational cost of subsequent MLIP MD simulations. In this study, we use a configurationally diverse beryllium dataset and quadratic spectral neighbor analysis potential. We demonstrate that joint optimization of energy *versus* force weights, training set selection strategies, and convergence settings of the *ab initio* reference simulations, as well as model complexity can lead to a significant reduction in the overall computational cost associated with training and evaluating MLIPs. This opens the door to computationally efficient generation of high-quality MLIPs for a range of applications which demand different accuracy *versus* training and evaluation cost trade-offs.

## 1 Introduction

Understanding atomic motion is fundamental for determining the physical and chemical properties of materials. Molecular dynamics (MD) simulations have been pivotal to address this challenge with applications ranging from drug design to nanotechnology.[1,2] Traditionally, MD simulation fell into one of two categories, either relying on empirical force fields to describe interatomic interactions that allow for long/large qualitative simulations with a cost that scales linearly with the number of atoms, or using *ab initio* quantum mechanical methods that enable small/short but very accurate simulations typically scaling cubically with the number of electrons. In the last decade, machine-learned interatomic potentials (MLIPs) have emerged as an alternative that promises near-quantum mechanical accuracy while scaling linearly with the number of atoms.[3–5]

[a]Theoretical Division T-1, Los Alamos National Laboratory, USA. E-mail: baghishov@utexas.edu; danny_perez@lanl.gov
[b]Oden Institute for Computational Engineering & Sciences, University of Texas at Austin, USA
[c]Department of Chemistry, University of Texas at Austin, USA
[d]Max Planck Institute for Sustainable Materials, Germany

The most recent developments in the field have prioritized improving the accuracy of MLIPs by incorporating complex atomistic descriptors and sophisticated machine learning models. While such models can now achieve remarkable accuracy, their training requires substantial amounts of high-fidelity *ab initio* training data and their evaluation can be thousands of times more expensive than traditional force fields, leading to significant computational costs at both training and evaluation times.[6–14] In contrast, other efforts prioritize applications such as high-throughput materials discovery, simulations of large atomic systems, or long timescale simulations, where minimizing the model's training and evaluation costs is paramount, even at the expense of a decrease in accuracy.[15–18] A prominent example of this philosophy is the development of ephemeral data derived potentials (EDDPs), which are lightweight potentials rapidly generated from moderately converged data specifically for demanding tasks such as high-throughput structure searching.[19–21] Finally, the development of "foundation" or "universal" models—highly complex MLIPs, often graph neural networks trained across vast chemical spaces[22–24]—raises questions about the continued need for optimizing application-specific potentials. However, these "universal" models often require fine-tuning for specific material systems to achieve high accuracy.[25–27] Crucially, fine-tuning preserves the high computational cost associated with the complex architectures of these "universal" models, which can be orders of magnitude greater than simpler alternative MLIPs like the linear Atomic Cluster Expansion (ACE).[25] Thus, for applications demanding both robustness and speed, tailoring less complex, optimized MLIPs remains crucial. This requires a systematic optimization of the application-specific cost/accuracy trade-off, considering the quality of the training set, the complexity of the model, and the training procedures, which is the central theme of this paper.

This paper explores the critical trade-off between accuracy and computational cost inherent in fitting and evaluating MLIPs. Fig. 1 conceptually maps the key factors that we investigate to navigate this trade-off, starting with the choice of MLIP complexity dictated by an application's needs in terms of number of simulations, simulation size and timescale with respect to the available computational budget. The optimization involves balancing the MLIP's predictive error (*e.g.*, energy and force RMSE) against the computational costs for constructing the training set and evaluating the MLIP. The computational cost for generating the density functional theory (DFT) training set is limited by the choice of convergence parameters such as plane wave energy cut-off (ENCUT) and *k*-point mesh sampling, as well as the total number of atomic configurations. In this paper, we refer to the ENCUT and *k*-point sampling parameters collectively as 'DFT convergence' or simply 'convergence'. The choice of convergence parameters dictates the numerical accuracy of DFT data relative to the fully converged basis set limit. Increasing ENCUT monotonically improves the absolute energy. This might suggest that it is only a constant shift to the potential energy surface, which would allow an MLIP to remain accurate up to a constant. However, the magnitude of this energy correction is structure-dependent. This variation introduces significant errors in the relative energies, with a magnitude comparable to *k*-point sampling errors (see Fig. S1 in the SI). Consequently, one cannot use an arbitrary ENCUT and assume that the MLIP will be accurate up to a constant; both convergence parameters must be investigated. It is important to distinguish this numerical convergence from the intrinsic accuracy in comparison to experiment dictated by the choice of the exchange–correlation functional, which is not considered here. Instead, the MLIP is benchmarked against the fully converged DFT results with a given functional.

While requiring tight DFT convergence settings is common,[28–30] it incurs substantial computational cost. We demonstrate that utilizing reduced-convergence DFT training sets can be sufficient provided the energy and force contributions are appropriately weighted during training. Furthermore, systematic sub-sampling techniques can identify the most informative configurations, drastically reducing the required training set size. By considering these aspects alongside the choice of MLIP complexity (which governs the computational cost of evaluation), we perform a joint Pareto analysis, conceptually illustrated by the optimal surface in Fig. 1. Our findings reveal that it is possible to achieve near-optimal MLIP accuracy with small, lower-convergence DFT training sets. This is especially true when using computationally efficient, reduced-complexity MLIPs. This underscores the substantial benefits of jointly optimizing model complexity, training set convergence, and training set to generate application-specific MLIPs with superior accuracy/cost characteristics.

The paper is organized as follows: Sec. 2 details the computational methodologies employed in this study. This includes the preparation of the DFT training set across six distinct convergence levels (Sec. 2.1), the description of the spectral neighbor analysis potential (SNAP) formalism and its quadratic extension (qSNAP), detailing variations in model complexity and the training procedure involving energy and force weighting (Sec. 2.2), and the leverage score technique utilized for efficient data sub-sampling (Sec. 2.3). Sec. 3 presents and discusses our findings. We first quantify the nature and magnitude of errors introduced by varying DFT convergence levels (Sec. 3.1), followed by analyzing how these DFT errors propagate into the trained MLIPs (Sec. 3.2). Next, we explore the effects of energy–force weighting (Sec. 3.3) and training set size (Sec. 3.4) on MLIP errors. Subsequently, we perform a multi-objective optimization, integrating DFT convergence, training set size (informed by leverage sampling), energy–force weighting and MLIP complexity to map out the Pareto front of accuracy *versus* computational cost (Sec. 3.5). Sec. 4 discusses the implications of these findings and quantifies the reduction in computational cost. Finally, Sec. 5 summarizes the key conclusions drawn from this work and discusses their implications for the fitting of application-specific MLIPs by inverting the parameter selection to tailor to specific application requirements for accuracy and computational cost.

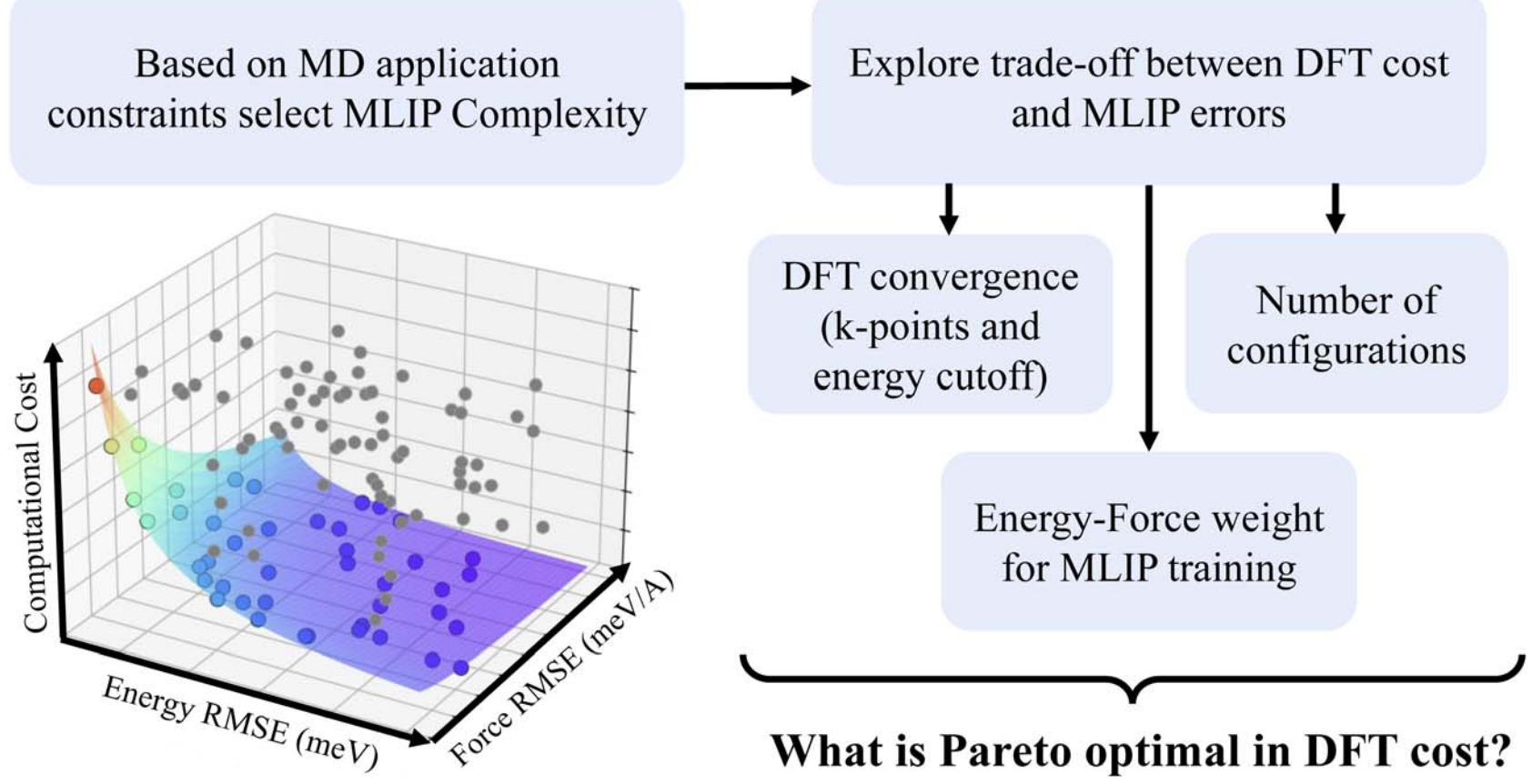

Fig. 1 Application-specific MLIPs are constructed starting from the desired application of the MLIP which restricts the computational costs per evaluation and consequently the complexity of the MLIP. This restricted MLIP complexity additionally impacts the benefit of increasing the training set size and *ab initio* convergence of the training set. Finally, we identify the energy *versus* force weight as a central parameter to position the potential on the Pareto front of computational cost, energy RMSE and force RMSE.

# 2 Methods

### 2.1 DFT training set

We generated a training and testing set of atomic beryllium configurations using the information entropy maximization approach introduced in ref. 31–33. This technique autonomously generates diverse datasets by creating atomic configurations that maximize the information entropy of the dataset in a feature space while bypassing the need for manual dataset curation by human experts. The MLIPs fitted to such diverse datasets were shown to be extremely robust and transferable.[32,33] We rely on a subset of 20 000 configurations selected from the dataset introduced in ref. 32, which was uniformly rescaled from the equilibrium lattice constant of tungsten to the equilibrium lattice constant of beryllium. This approach leverages the fact that both systems are unary, allowing the reuse of diverse geometric configurations from the entropy maximization algorithm without repeating the generation step for beryllium. Each configuration contains on average 50 atoms. When training MLIPs, the dataset is split evenly into a training and a testing set of 10 000 configurations each. While the system is chemically simple, the entropy-maximization method generates extreme topological diversity, even compared to datasets enriched by active learning.[32,33] The extremely broad coverage of the feature space leads to ultra-robust MLIPs, in contrast to models trained on conventional datasets that routinely dramatically fail at predicting the properties of entropy-maximized configurations. This diversity makes the expressivity tradeoffs of the MLIPs especially pronounced.

Table 1 Six levels of DFT convergence are introduced, referenced as convergence level 1 to 6, with different $k$-point mesh samplings and plane-wave energy cut-offs, resulting in different averaged evaluation run times per configuration

| Convergence level | $k$-point spacing, $Å^{-1}$ | Energy cut-off, eV | Average run time per configuration, sec |
|---|---|---|---|
| 1 | Gamma point only | 300 | 8.33 |
| 2 | 1.00 | 300 | 10.02 |
| 3 | 0.75 | 400 | 14.80 |
| 4 | 0.50 | 500 | 19.18 |
| 5 | 0.25 | 700 | 91.99 |
| 6 | 0.10 | 900 | 996.14 |

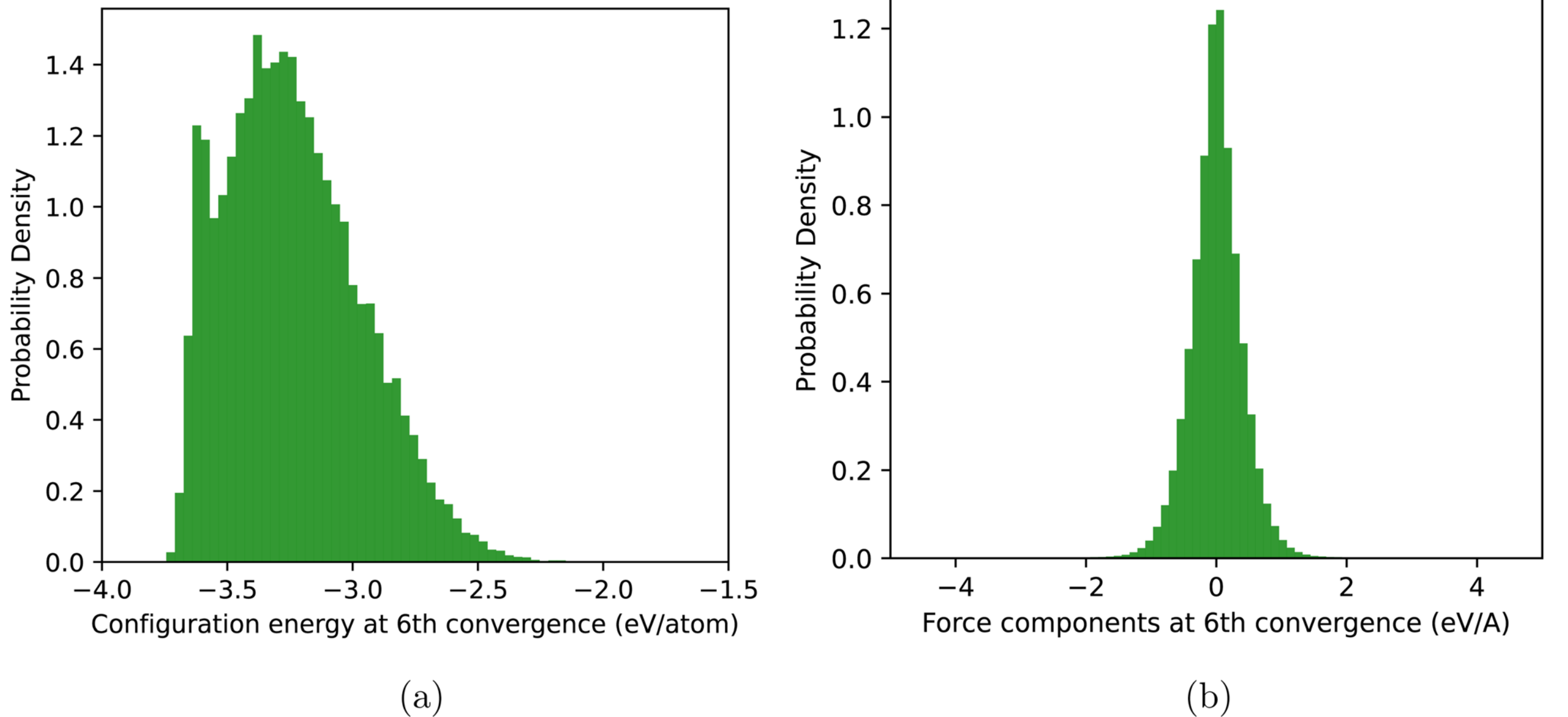

Fig. 2 Probability density of configuration energies and forces in the entropy maximized dataset for beryllium evaluated at convergence level 6 (highest). (a) Distribution of per-atom energies. (b) Distribution of force components.

Reference energies and forces are calculated using pyiron workflow manager[34] and the Vienna *Ab initio* Simulation Package (VASP)[35,36] at six levels of DFT convergence, as shown in Table 1. The calculations employed the projector augmented-wave (PAW) method in conjunction with the Perdew–Burke–Ernzerhof (PBE) exchange–correlation functional. To model electronic smearing, we utilized the Methfessel–Paxton scheme (ISMEAR = 1) with a smearing parameter of SIGMA = 0.2 eV. The table also reports the average simulation time for single point evaluations using a single NVIDIA A100 GPU; the computational effort required to generate the reference data is seen to vary by a factor of around 100 between low (level 1) and high (level 6) convergence simulations.

The per-atom energy distribution (Fig. 2a) spans from −3.75 to −2 eV per atom. Although this dataset is challenging to fit due to its diversity, it has been shown to produce highly transferable and robust potentials. The force component distribution (Fig. 2b) shows a standard deviation of 0.44 eV Å$^{-1}$ and ranges from −10 to 10 eV Å$^{-1}$, with most values concentrated between −2 and 2 eV Å$^{-1}$.

### 2.2 ML potential

The spectral neighbor analysis potential (SNAP) formalism expands atomic energies and forces using linear combinations of bispectrum components derived from 4D spherical harmonics, providing rotationally invariant descriptors of local atomic environments.[7] The bispectrum components systematically capture the geometry of an atom's local neighborhood by projecting its neighbor density function onto a basis of hyperspherical harmonics. This creates a set of descriptive coefficients that are invariant to rotations, translations, and permutations of identical atoms, making them robust descriptors for atomic potentials. In the following, we employ its quadratic extension, qSNAP, which incorporates quadratic bispectrum terms, improving the accuracy for complex bonding environments while maintaining computational efficiency.[37] A key advantage of the SNAP/qSNAP formalism is that the potential is linear with respect to the bispectrum descriptors. This casts the fitting process as a simple linear regression, which is computationally inexpensive compared to training non-linear models like neural networks, allowing for extensive studies that requires training thousands of models. qSNAP descriptors were obtained using FitSNAP software.[38] It is important to note that the approaches described below are general and can be applied to other functional MLIP forms. In the present context, the use of qSNAP allowed for an extensive exploration of the tradeoffs explored in this study due to the low cost of training each model. The atomic reference energy for beryllium is not fixed beforehand but is treated as a free parameter determined implicitly as part of the linear least-squares fitting process.

To study the benefits of increased DFT convergence for MLIPs with varying levels of complexity, we control the MLIP complexity for the qSNAP MLIP *via* the $2J_{\max}$ parameter, which directly controls the angular order of the spherical harmonic expansion. Conceptually, $2J_{\max}$ determines the angular resolution of the descriptors; a larger value allows the model to capture finer details and more complex geometric arrangements in the local atomic environment. Table 2 reports how increasing $2J_{\max}$ affects the number of descriptors, which corresponds to the number of MLIP coefficients and the computational cost of evaluation for applications, like calculating MD trajectories with the MLIP. Notably, the evaluation cost varies by nearly an order of magnitude between $2J_{\max} = 4$ and 10, highlighting the direct impact of model complexity on computational cost for the application of the MLIP. While other hyperparameters, such as the radial cutoff and element-specific weights, are important for defining the scope of the local environment, $2J_{\max}$ is the principal hyperparameter that governs the flexibility and descriptive power of the qSNAP functional form itself, and therefore serves as the primary control for model complexity.

Table 2 Number of bispectrum components (descriptors) and computational speed for increasing values of $2J_{\max}$. Taken from Wood and Thompson, 2018 (ref. 37)

| $2J_{\max}$ | Number of descriptors | Atoms-timestep per second |
|---|---|---|
| 4 | 105 | $4 \times 10^5$ |
| 6 | 465 | $1 \times 10^5$ |
| 8 | 1540 | $4 \times 10^4$ |
| 10 | 4186 | $2 \times 10^4$ |

In spite of what its name suggests, training a qSNAP MLIP can be cast as a linear regression task, which greatly facilitates the training of the large number of models which are presented below. In the following, for simplicity we consider only regression to reference energies and forces, although other quantities such as stresses can be added.

The training of the linear model is based on minimizing a weighted least squares loss function:

$$L = \sum_{m=1}^{M} \left\{ \frac{{w_{\mathrm{E}}}^2 \left( \hat{E}_m - E_m \right)^2}{{N_m}^2} + \sum_{i=1}^{3N_m} {w_{\mathrm{F}}}^2 \left( \hat{F}_{mi} - F_{mi} \right)^2 \right\} \tag{1}$$

where $M$ is the number of configurations, $m$ indexes a particular configuration, $N_m$ represents the number of atoms in configuration $m$, and $i$ refers to an atomic force component. The reference energies and forces are denoted as $E_m$ and $F_{mi}$, while their predicted counterparts are $\hat{E}_m$ and $Fmi$, respectively. The terms $w_{\mathrm{E}}$ and $w_{\mathrm{F}}$ are weights assigned to energy and force contributions, respectively.

In matrix form, the solution to this minimization problem is found by solving the weighted least squares equation:

$$Wy = WX\beta \tag{2}$$

where $X \in \mathbb{R}^{n \times p}$ is the descriptor matrix with $n$ representing the total number of energy and force components in the dataset, and $p$ being the number of qSNAP descriptor components, $y \in \mathbb{R}^n$ is the vector of reference values (including both energies and forces), and $\beta \in \mathbb{R}^p$ is the vector of MLIP coefficients. The diagonal weight matrix $W \in \mathbb{R}^{n \times n}$ shown in eqn (1) provides a user-adjustable relative weighting of energy and force terms in the loss function.

$$W = \begin{bmatrix} \frac{w_E}{N_1} & 0 & 0 & 0 & \cdots & 0 \\ 0 & w_F & 0 & 0 & \cdots & 0 \\ 0 & 0 & w_F & 0 & \cdots & 0 \\ 0 & 0 & 0 & \frac{w_E}{N_2} & \cdots & 0 \\ \vdots & \vdots & \vdots & \vdots & \ddots & \vdots \\ 0 & 0 & 0 & 0 & \cdots & w_F \end{bmatrix} \quad (3)$$

This formulation ensures that minimizing the weighted least squares problem is equivalent to minimizing the loss function defined in eqn (1). It will be shown below that the choice of weights $w_E$ and $w_F$ plays a critical role in balancing the influence of energy and force errors and depends on both the complexity of the MLIP model and the DFT convergence of the training set (Sec. 3.3).

### 2.3 Training set sub-sampling with leverage score

A key aspect of the MLIP design challenge is the curation of the training set. Indeed, depending on the complexity of the MLIP, less than 10 000 configurations could be sufficient to obtain a converged MLIP even when using a high-diversity dataset.[31–33] This begs the question of how to choose a proper subset of configurations to evaluate with DFT. In the following, we use a leverage score based strategy. Leverage quantifies how much a configuration's features in descriptor space deviate from the average, allowing us to identify configurations with distinctive features. It can also be interpreted in terms of the sensitivity of the $i$th predicted value $\hat{y_i}$ on the $i$th dependent value $y_i$ (where the $y$s can be either energies of a particular configurations or force components of a particular atom). High-leverage points therefore have the potential to significantly affect predictions carried out with the trained MLIP. This concept is closely related to the maximum volume approach or D-optimality criterion used in active learning, which selects data points that expand the coverage in descriptor space.[39] Our sub-sampling procedure consists of randomly sampling configurations from the 10 000 candidates with probabilities proportional to their leverage scores until a subset of the desired size is obtained.

The leverage score of data item $i$ is the corresponding diagonal element of the so-called hat matrix $H = X(X^TX)^{-1}X^T$. A numerically stable and efficient procedure to evaluate the leverage involves the singular value decomposition (SVD) of $X = U\Sigma V^T$. From basic properties of the SVD, it can easily be shown that:

$$\begin{aligned} H &= X(X^TX)^{-1}X^T = U\Sigma V^T(V\Sigma U^T U\Sigma V^T)^{-1}V\Sigma U^T \\ &= U\Sigma V^T(V\Sigma^2 V^T)^{-1}V\Sigma U^T = U\Sigma V^T V^T \Sigma^{-2} V^T V\Sigma U^T \\ &= U\Sigma V^T V\Sigma^{-2} V^T V\Sigma U^T = UU^T \end{aligned} \quad (4)$$

Since each configuration corresponds to a block of rows in the descriptor matrix $X$ (one for total energy and $3N_m$ for atomic forces), we explore two strategies to assign a single score to each configuration. The first, which we call 'regular leverage sampling', uses the leverage score calculated from the energy descriptor row only, a method analogous to CUR decomposition.[40] The second, 'block leverage sampling', calculates a total score by summing the individual leverage scores of all rows (energy and all force components) associated with that configuration, an approach analogous to block CUR decomposition.[41]

## 3 Results and discussion

### 3.1 Quantifying errors in low-convergence DFT calculations

Understanding the errors introduced by different levels of DFT convergence requires recognizing the distinct convergence behaviors of the primary convergence parameters: the plane-wave energy cutoff (ENCUT) exhibits monotonic energy decrease governed by the variational principle, whereas $k$-point sampling, approximating Brillouin zone integration, shows generally non-monotonic oscillatory convergence. In our study we analyze ENCUT and $k$-point spacing together as a measure of DFT convergence. While ENCUT convergence is monotonic, its effect on total energy is not uniform across different atomic configurations. As demonstrated by our convergence tests (see Fig. S1 in the SI, the change in energy from increasing ENCUT varies by up to 5 meV per atom between structures. This differential convergence is on a similar scale as $k$-point sampling errors and directly impacts the relative energies that an MLIP learns. Therefore, treating both ENCUT and $k$-points as components of the overall DFT convergence is essential for accurately assessing the cost-benefit trade-offs in MLIP training. We characterize the statistical properties of these errors observed in our dataset, providing a guide for the development of the fitting strategies introduced below. Fig. 3 compares DFT calculations at the 2nd and 6th convergence levels, showing energy relationships (Fig. 3a) and force relationships (Fig. 3b). Both quantities show different statistical behaviors. The distribution of force errors is centered at zero and symmetric, suggesting that low-convergence forces could potentially be considered as "noisy" versions of exact forces, to a first approximation. Points in Fig. 3 are color-coded by the shortest cell vector length, highlighting that energy errors appear to be significantly affected by the largest $k$-point spacing along any cell dimension, suggesting that insufficient coverage in the Brillouin zone leads to systematic errors. Force errors don't show the same bias as energies although they are broader for cells which are shorter in at least one direction, but the effect is much smaller. These trends persist at other convergence levels, although the errors rapidly become smaller as convergence increases, as shown in SI Fig. S2 and S3. In addition, convergence level 1 (lowest) energies show a systematic overestimation relative to level 6 (highest) (Fig. 2). This strong bias is much less pronounced for convergence levels 2, 3, 4, and 5, where the energy errors are substantially smaller.

Error metrics are summarized in Table 3, which presents the root mean squared differences (RMSD) between the 6th convergence level and other convergence levels for the energies and forces. The "Unshifted" column reports raw errors. It is sometimes argued that errors between different DFT convergence levels can be resolved by a simple constant energy shift, such as referencing the bulk ground state, a consideration motivated by the faster convergence of energy differences due to empirically-observed error cancellation.[42] To test this hypothesis, the "Bulk shifted" column reports the RMSD after applying such a shift. The results show that this procedure is often detrimental, increasing the error for most convergence levels. This demonstrates that for a configurationally diverse dataset like ours, which includes many high-energy structures, a bulk reference point is insufficient to correct for convergence-related errors across the entire potential energy surface. The "Mean shifted" column, although a dataset-dependent measure, is included to distinguish a systematic energy offset from other sources of error. Finally, we observe that convergence level 5 is required to achieve energy errors lower than 1 meV per atom, which is often regarded as a target for accurate MLIPs.

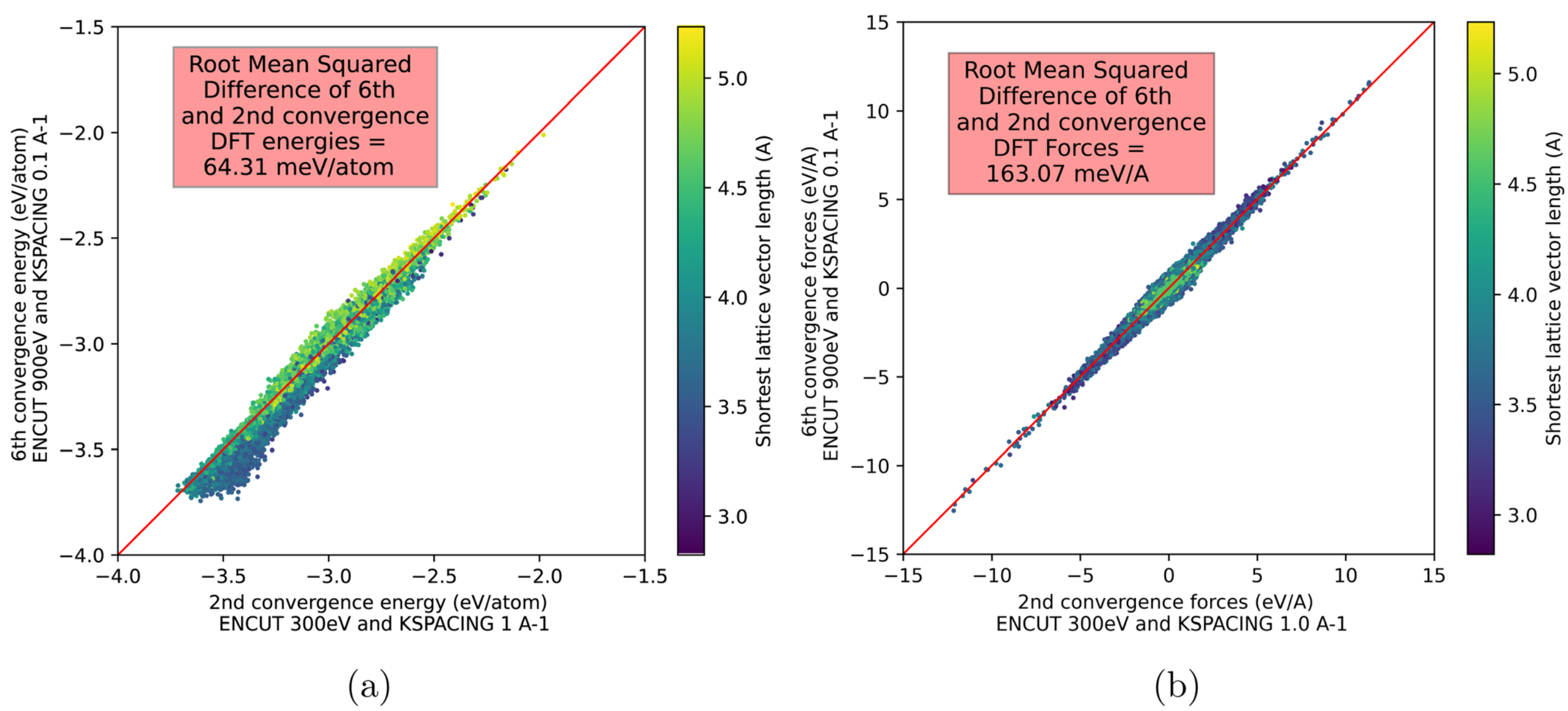

Fig. 3 Pairwise relationships between 2nd and 6th convergence level DFT data. (a) Energies and (b) forces.

Table 3 Energy and force root mean squared differences relative to the 6th convergence level

| | Energy, meV per atom | | | |
|---|---|---|---|---|
| Convergence level | Unshifted | Bulk shifted | Mean shifted | Forces meV Å$^{-1}$ |
| 1 | 497.05 | 437.96 | 221.72 | 417.76 |
| 2 | 64.31 | 192.14 | 47.46 | 163.07 |
| 3 | 15.34 | 20.21 | 15.28 | 77.64 |
| 4 | 5.59 | 59.34 | 5.39 | 51.17 |
| 5 | 0.54 | 2.42 | 0.50 | 10.17 |

We note that an analysis in terms of pointwise averages omits very important properties of energy and force errors that differentiates them from statistical noise. First, actual errors are correlated in that small changes in the positions of atoms are likely to incur similar errors, leading to smooth local distortions with respect to a fully converged potential energy surface. Second, errors can also be discontinuous, *e.g.*, when the *k*-point mesh discretely changes as a simulation cell is smoothly distorted. As will be shown in the next section, the different statistical properties of the energy and force errors can be used to mitigate their impact and reduce error propagation into MLIP models.

### 3.2 Effect of DFT convergence errors on MLIP training

To investigate the impact of errors introduced by lower-convergence DFT simulation on the accuracy of MLIPs, we trained qSNAP potentials to half of the datasets (10 000 configurations) evaluated at the six convergence levels and subsequently tested the models on the other half, using both the corresponding convergence level or the highest convergence level. In addition, the effect of the relative weight of energies and forces is explored. Unless otherwise noted, all energy RMSE values reported from this point forward are the raw 'unshifted' errors.

Table 4 and 5 report the root mean squared errors (RMSE) values for energy and force errors for MLIPs trained at different convergence levels, with both higher energy weight ($w_E$:$w_F$ =

Table 4 Energy root mean squared errors (meV per atom) for MLIPs trained on 10 000 configurations at different DFT convergence levels and with the highest level of MLIP expressivity ($2J_{max} = 10$)

| | Higher energy weight | | | | Higher force weight | | | |
|---|---|---|---|---|---|---|---|---|
| | | | Testing on 6th | Training | | | Testing on 6th | Training |
| Conv. Level | Training | Testing on self | Unshifted | Shifted | Training | Testing on self | Unshifted | Shifted |
| 1 | 111.74 | 117.91 | 473.42 | 164.02 | 187.84 | 188.55 | 458.17 | 104.85 |
| 2 | 30.90 | 33.18 | 53.76 | 32.17 | 39.26 | 40.06 | 50.21 | 25.85 |
| 3 | 11.01 | 11.67 | 11.08 | 10.99 | 15.12 | 15.48 | 9.24 | 9.14 |
| 4 | 6.20 | 6.58 | 5.80 | 5.62 | 8.84 | 9.01 | 8.16 | 8.04 |
| 5 | 4.79 | 5.06 | 5.05 | 5.05 | 7.70 | 7.84 | 7.85 | 7.85 |
| 6 | 4.77 | 5.05 | 5.05 | 5.05 | 7.70 | 7.84 | 7.84 | 7.84 |

Table 5 Force root mean squared errors (meV Å$^{-1}$) for MLIPs trained on 10 000 configurations at different DFT and with the highest level of complexity ($2J_{max} = 10$). No shift correction is applied to force errors

| | Higher energy weight | | | Higher force weight | | |
|---|---|---|---|---|---|---|
| Prec. Level | Training | Testing on self | Training | Testing on self | Training | Testing on self |
| 1 | 931.09 | 944.15 | 900.29 | 375.83 | 379.2 | 242.75 |
| 2 | 235.72 | 238.89 | 186.67 | 189.33 | 190.69 | 116.66 |
| 3 | 143.30 | 145.03 | 124.53 | 130.66 | 132.06 | 108.57 |
| 4 | 120.46 | 121.86 | 111.50 | 116.4 | 117.67 | 106.89 |
| 5 | 109.41 | 110.87 | 110.51 | 105.71 | 106.98 | 106.6 |
| 6 | 109.04 | 110.52 | 110.52 | 105.32 | 106.6 | 106.6 |

150 : 1) and higher force weight ($w_E$:$w_F$ = 12 : 1). Note that the datasets contain around 150 times more force components than energies as each configuration contains on average 50 atoms (see Sec. 2.1). Each row in the table corresponds to the convergence level of the training set. The "Testing on self" column reports the RMSE values when tested on the same convergence level as the training set, while the "Testing on 6th" column reports errors when tested on 6th (highest) convergence level data. The ultimate goal is to obtain models with low errors when tested against the highest convergence level. Testing errors are evaluated using both an unshifted potential and a shifted potential, where the latter includes a constant energy offset equal to the mean prediction error. Table 5 presents similar data for force RMSE values, with the exception of the absence of a shift correction.

Table 4 highlights that training errors are larger for MLIPs trained on lower convergence data compared to higher convergence, indicating that the potential energy surface becomes smoother and hence easier to fit as the convergence increases. This can be related to the discussion above (see Sec. 2.1) where the impact of discontinuities and inconsistencies due to incompatible *k*-point meshes and limited plane wave energy cut-offs is expected to decrease with increasing convergence, producing smoother and more internally consistent potential energy surfaces, as commonly reported in the literature.[42,43] Interestingly, we observe that testing errors are generally lower when models trained on lower convergence data are tested on 6th convergence data than on their own level of convergence. This observation suggests that the inability of MLIPs to capture unphysical behavior such as energy discontinuities/inconsistencies due to discrete changes in *k*-point mesh sampling can, in fact, be an advantage, since it can be used to partially recover the behavior of smoother high-convergence data. Supporting this interpretation, SI Fig. S4–S7 show strong correlations between energy and force residuals from MLIPs trained on low-convergence data and actual DFT error between high- and low-convergence DFT energies and forces, indicating that artifacts in the low-convergence DFT energy surface are indeed partially corrected by the MLIP. This trend is especially evident when large force weights are used during training, as reflected in Table 4, where the lowest energy RMSE values for the 1st, 2nd, and 3rd convergence levels occur at higher force weight. However, this is no longer true for the 4th and 5th convergence levels, where DFT errors are small and the MLIP errors become limited by the model's intrinsic ability to capture the full complexity of the dataset, as indicated by the error saturation when training to higher DFT convergence data.

### 3.3 Energy–force weight dependence

Fig. 4 illustrates how energy and force RMSEs are affected by energy *versus* force weights. In these plots, the relative weight of energies *versus* forces in the regression increases/decreases from the top left to the bottom right. The dotted line represents the RMSE when evaluated at the same convergence level as the training and testing set, while the solid line shows errors evaluated on the testing set evaluated at the highest (6th) convergence level.

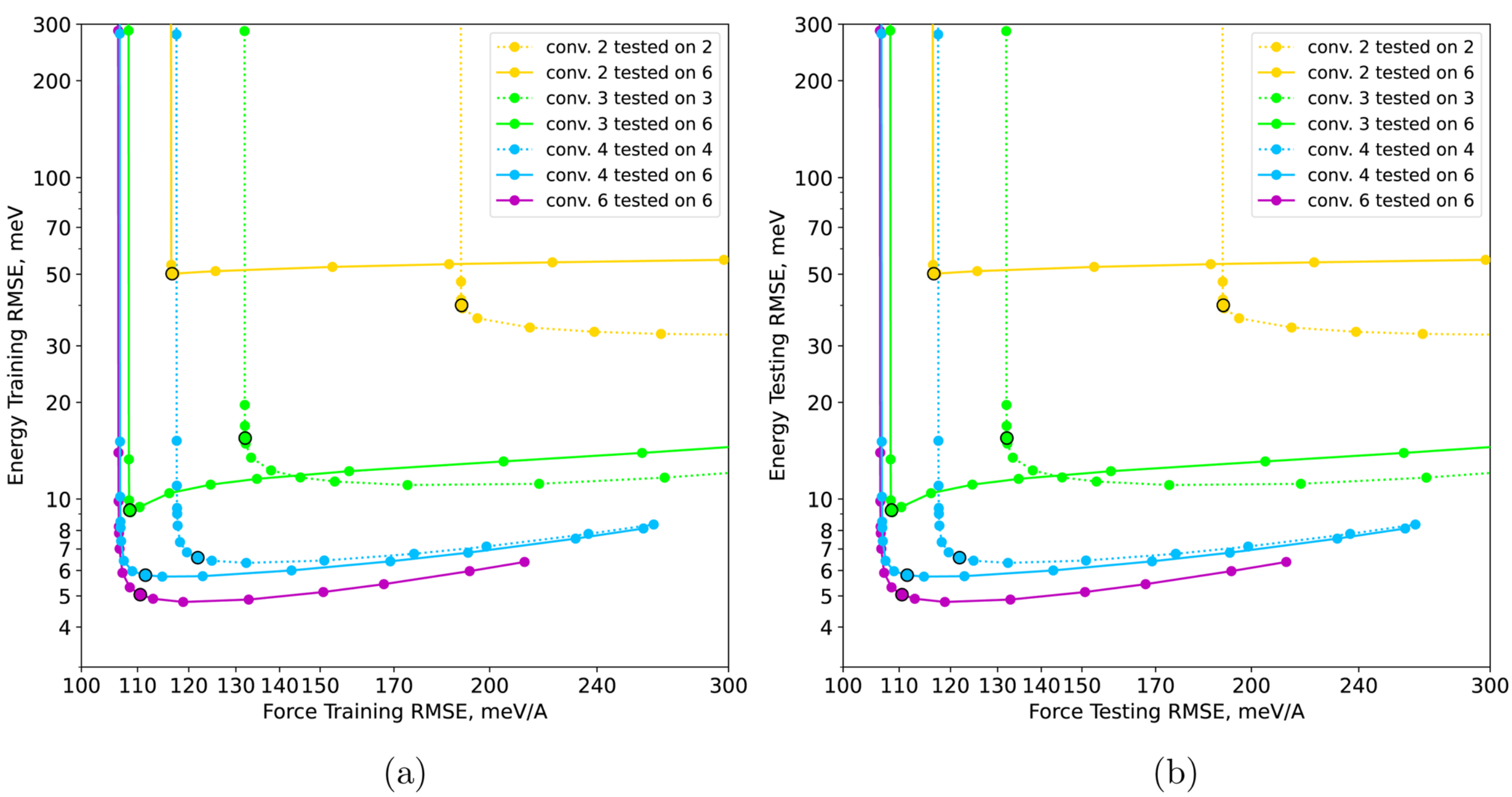

Fig. 4 Pareto front of energy and force testing errors for different energy *versus* force weights in qSNAP fitting ($2J_{max} = 10$) across various convergence levels trained on 10 000 configurations. Convergence level 5 is omitted as it visually overlaps with convergence level 6. Convergence level 1 data fall out of the range of these plots. (a) Training errors and (b) testing errors.

Perhaps counterintuitively, increasing the energy weight of convergence levels 2 and 3 leads to higher training and testing energy errors when measured at the 6th convergence level (solid line). In these cases, the Pareto front almost collapses to a single point which simultaneously provides the lowest energy and force errors. This outcome stems from the nature of errors in low-convergence DFT calculations, caused by insufficient sampling of the *k*-point mesh and sharp features in the potential energy surface based on the restricted number of plane waves, affecting the total energy of the supercell. The forces are less sensitive and exhibit faster convergence compared to energies with respect to the DFT convergence parameters like plane wave cut-off and *k*-point density.[43] This is also illustrated by the force error in Fig. 3b which is more symmetric and Gaussian-like, suggesting force-related errors less effected by a reduced plane wave energy cut-off or a reduced *k*-point mesh sampling, resulting in a statistically more well-behaved training set for learning compared to the errors in energy. Consequently, focusing training excessively on low-convergence energies by increasing energy weight can lead the MLIP to partially learn the incorrect low-convergence potential energy surface, resulting in these systematic errors that distort the potential energy surface away from the high-convergence reference. In contrast, by focusing more on forces, the MLIP can most efficiently "average-out" errors, effectively learning a smoother representation of the low-convergence energy landscape which is closer to the fully converged high-convergence potential energy surface. This suggests that leveraging the larger, statistically more robust force dataset *via* increased weighting can mitigate the impact of low-convergence energy noise. Therefore, carefully adjusting the relative weighting of energies and forces is crucial when training MLIPs on lower-convergence data, as increasing force weights can, perhaps paradoxically, produce MLIPs yielding better energies in comparison to the high-convergence potential energy surface as well as better force convergence. A similar principle has been demonstrated in the context of multi-fidelity learning, which shows that forces from a lower-accuracy level of theory can be effectively combined with higher-accuracy energies to produce robust potentials.[44] Note, the increase in test energy errors at high DFT convergence (levels 4 and 6, right panel of Fig. 4) can be attributed to the model using its fitting freedom to overfit to the small number of energy data points when they are weighted higher. This behavior is closely related to model misspecification; the model's flexibility, including adjusting the implicit atomic reference energy, is used to capture specific features of the training set energies at the expense of generalizability. This overfitting can be regularized by increasing the force weight.

In most applications that require simulating material properties that are not accessible with direct DFT simulations, the complexity of the MLIP is limited by the available computational resources *e.g.*, the complexity of the MLIP is chosen based on the goal to achieve a fixed target in terms of the number MD simulation time steps required to investigate a given physical phenomenon of interest. The results presented in the previous section suggest that the impact of DFT errors varies based on the complexity of the MLIP, which provides the opportunity to reduce the DFT convergence and increase the computational efficiency during the dataset curation. Indeed, highly complex MLIPs are expected to be more prone to learning the DFT errors compared to simpler models, and so will intrinsically require higher convergence DFT training sets, while simpler models benefit from lower-convergence training sets, reducing the computational cost to construct the DFT training set. To explore this, the previous analysis is repeated for different model complexities. For qSNAP potentials, this is achieved by varying the angular order of the bispectrum expansion, which is commonly referred as the $2J_{max}$ parameter[7] (see Sec. 2.2). As shown in Table 2, increasing $2J_{max}$ corresponds to a rapid increase in the number of coefficients in the MLIP, and hence to an increase in complexity.

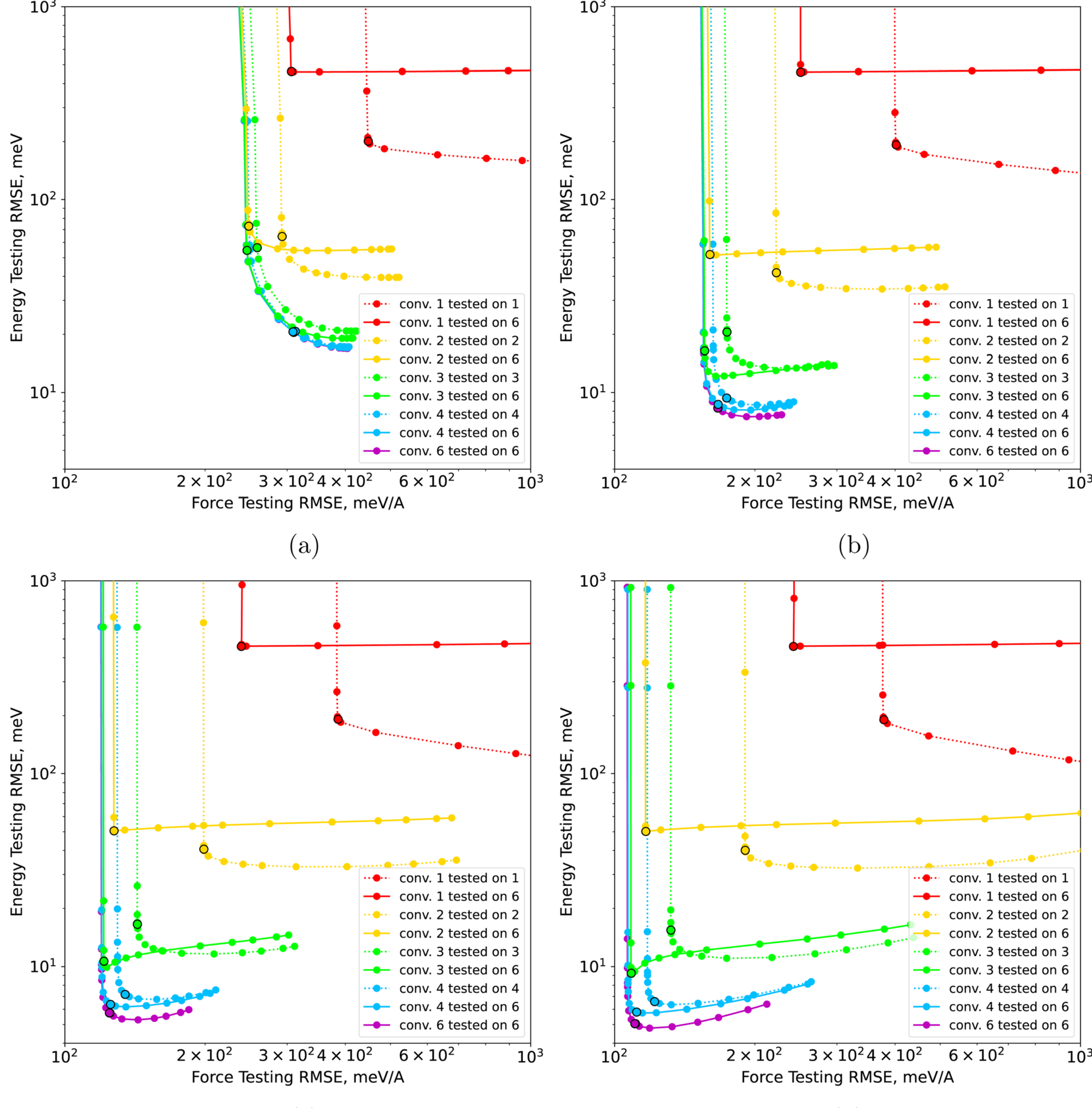

Fig. 5 Energy/force Pareto front of testing errors with different energy *versus* force weights in the fitting and at different convergence levels for various $2J_{max}$ values fitted to a training set with 10 000 configurations. (a) $2J_{max} = 4$, (b) $2J_{max} = 6$, (c) $2J_{max} = 8$, and (d) $2J_{max} = 10$.

Fig. 5 shows that simple MLIPs are indeed less sensitive to the errors in low-convergence DFT training sets. For example, for $2J_{max} = 4$, the potentials trained on convergence levels 3, 4, and 5 yield similar errors when tested on a convergence level 6 testing set. In contrast more complex MLIPs with $2J_{max} = 10$ show a more pronounced dependence on the training set convergence level. This suggests that lower complexity MLIPs can leverage low-convergence DFT training sets compared to higher complexity MLIPs, as they are less able to learn spurious features of the low-convergence DFT potential energy surface.

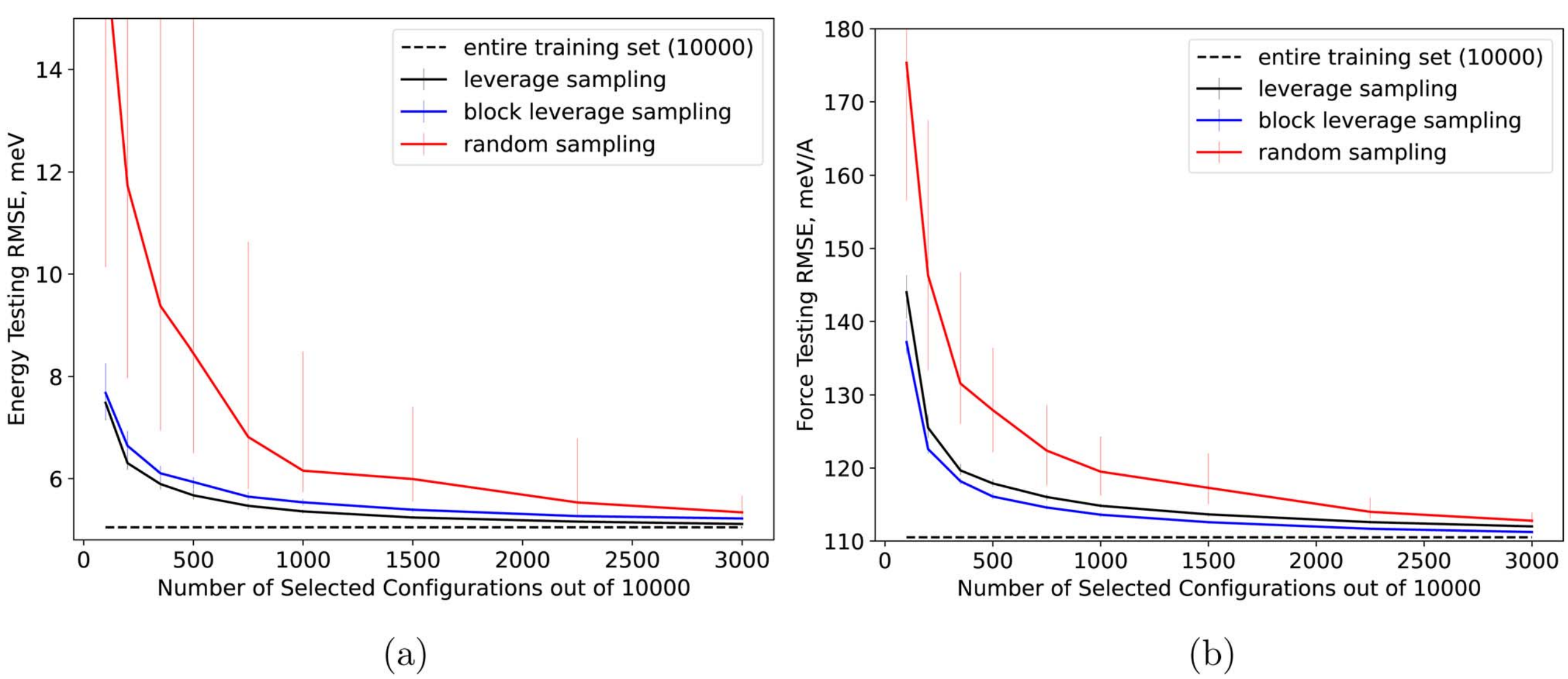

Fig. 6 Comparison of regular and block leverage sampling with random sampling ($2J_{max} = 10$) for the highest level of DFT convergence and a fixed energy *versus* forces weight ($w_E/w_F = 150$). (a) Energies and (b) forces.

### 3.4 Data sub-sampling *via* leverage score

A critical factor in efficiently curating training sets for MLIPs is the trade-off between the number of configurations and the convergence of the underlying DFT simulations to evaluate the configurations. Here, we explore the effect of data sub-sampling strategies on the accuracy of the obtained MLIP. In this respect, Fig. 6 highlights that leverage sampling significantly outperforms random sampling, in terms of decay rate of both the energy and force errors and of their variance with increasing number of configurations. Significant savings of computational cost are achieved, with only a few hundred configurations being required to consistently reach within 1 meV per atom of the result obtained when using all 10 000 configurations where the error is dominated by the finite expressivity of the MLIP. This results in a reduction of computational cost by a factor of 10 compared to random sampling. Similarly, MLIPs can approach the limiting force errors by about 10 meV $Å^{-1}$ using 3 to 4 times less configuration than required by random selection. Finally, block leverage sampling is observed to yield lower force errors while regular leverage sampling leads to lower energy errors, although the differences between the two approaches are modest. It is important to note that the computational cost of leverage sampling is minimal, as it can be obtained at a computational cost comparable to that of a single linear regression solution on the whole dataset of 20 000 configurations. Importantly, configurations can be prioritized using leverage sampling without the need to carry out the related DFT simulation first, as only the features of each atomic configuration are required to compute the leverage score, but the energies and forces computed by DFT are not required.

### 3.5 Multi-objective optimization of application-specific MLIPs

To understand the combined influence of DFT convergence, training set size, energy–force weighting, and MLIP complexity ($2J_{max}$) on the cost/accuracy trade-off, we performed a systematic joint exploration. We trained numerous qSNAP potentials using a full factorial design, varying convergence levels (1–6), subset sizes selected *via* leverage sampling, energy–force weights, and four $2J_{max}$ values. This comprehensive analysis allows us to map the Pareto-optimal front of possible MLIPs relating DFT computation time (cost) to energy and force RMSE (accuracy). Fig. 7 reports the MLIPs on this Pareto front, highlighting the broad families of optimal MLIPs in this multi-objective setting. In the figure, marker shapes encode training set convergence levels while their colors denote the total DFT computation time required to obtain the training set. The markers align along rough lines that correspond to varying training subset sizes at a specific energy–force weight. Rather than pinpointing a specific MLIP from the force–energy Pareto front, we analyze the front as a whole, offering insights into the trade-offs between DFT convergence and MLIP settings depending on the desired accuracy in energy *vs.* forces, which is a user-specified preference. MLIPs that are not Pareto optimal are hidden.

Several key conclusions emerge: neither the 1st nor 2nd convergence levels appear on the Pareto front, indicating that smaller higher-convergence subsets always outperform larger very-low-convergence datasets. We postulate that this reflects an inherent trade-off where larger number of configurations of very noisy data are required to “average out” intrinsic errors, compared to high-convergence data where the ultimate accuracy limit-the point at which errors are controlled only by the model's finite complexity can be expected to occur earlier. This phenomenon can be frequently observed on the Pareto front (especially in panels (a) and (b) of Fig. 7) where different DFT convergence levels can lead to similar accuracies and overall DFT computational costs, indicating that the lower-convergence training sets were hence larger than their high-convergence counterparts. Of course, whether the extra amount of low-convergence training set configuration can be obtained at a sufficiently low computational cost compared to a smaller number of configurations at higher DFT convergence is application-specific. Similarly, the 6th (highest) level convergence training set is mostly absent except at the very edge of the accessible error range, due to a marginal increase in convergence obtained in spite of the significantly higher computational cost.

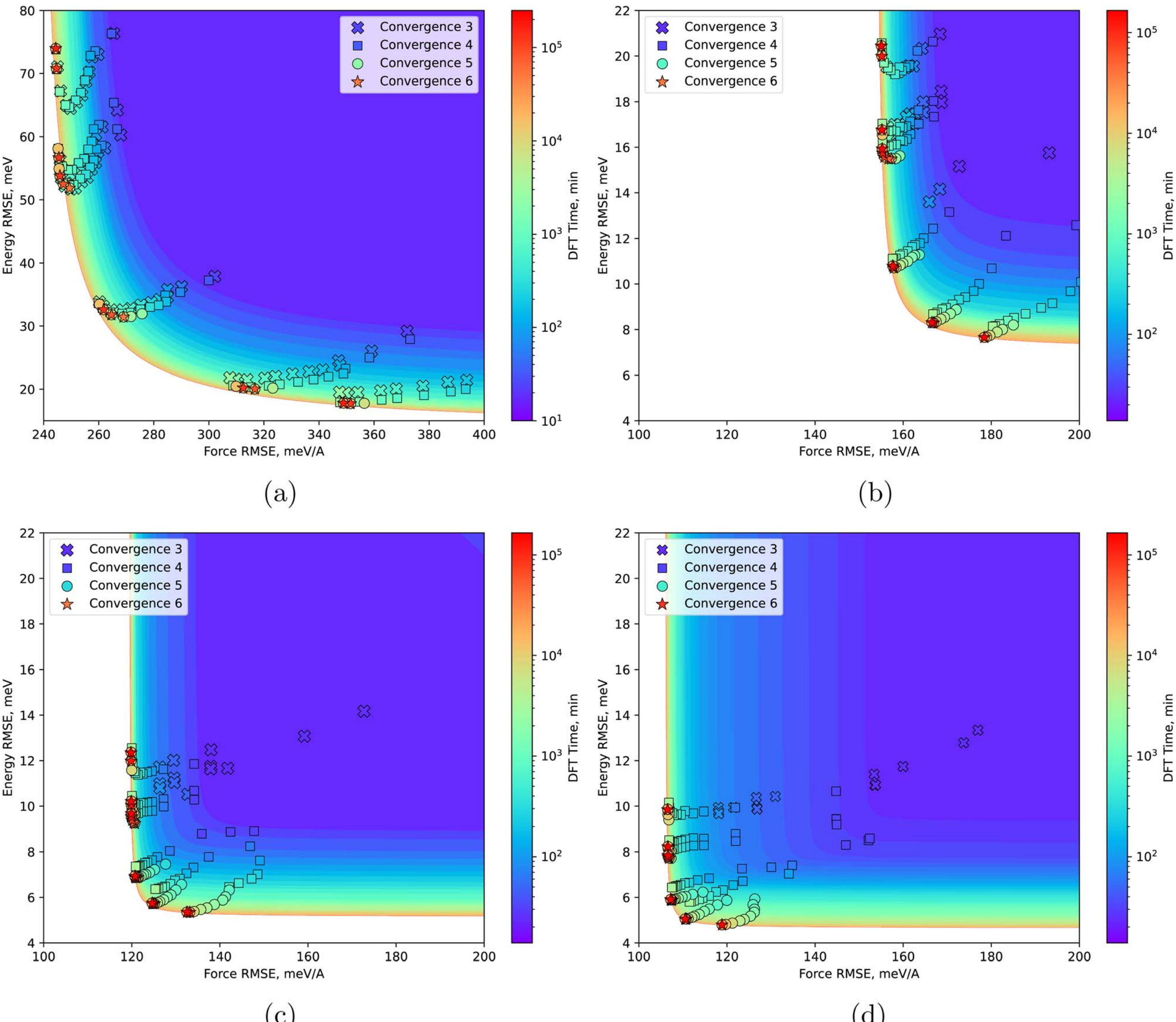

Fig. 7 Testing RMSEs (energy *vs.* force) for Pareto-optimal MLIPs of varying complexity ($2J_{max} \in \{4, 6, 8, 10\}$); all models were tested on level 6 convergence DFT testing sets. Marker color and approximate background surface indicate the computational cost of the DFT simulation for all MLIPs. Marker shapes distinguish the level of DFT convergence used for the training set. Markers are organized into lines, where each line corresponds to a specific energy–force weight ($w_E/w_F \in \{5, 10, 12.25, 50, 150, 300\}$). Along each line, individual markers denote different training set sizes, ranging from 100 to 10 000 configurations. (a) $2J_{max} = 4$, (b) $2J_{max} = 6$, (c) $2J_{max} = 8$, and (d) $2J_{max} = 10$.

Correlating these findings with Table 3 suggests that different levels of DFT convergence corresponding to the error much lower than the ultimate accuracy achievable by the MLIP due to finite complexity are unlikely to be optimal, as the amount of extra information gained by high-convergence DFT simulation has a limited impact on the accuracy of the MLIP. This explains the scarcity of DFT simulation of level 6 convergence on the Pareto front. Similarly, the value of low-convergence DFT simulation is limited when the DFT errors significantly exceed the accuracy achievable by the MLIP due to the need to average out these errors. This is also consistent with the absence of the 1st and 2nd convergence levels of DFT convergence on the Pareto front, as their intrinsic errors exceed the representation capabilities of all MLIPs considered here. These observations are consistent with a rule of thumb, where matching the ultimate accuracy of the MLIP and the convergence of the DFT simulation is desirable.

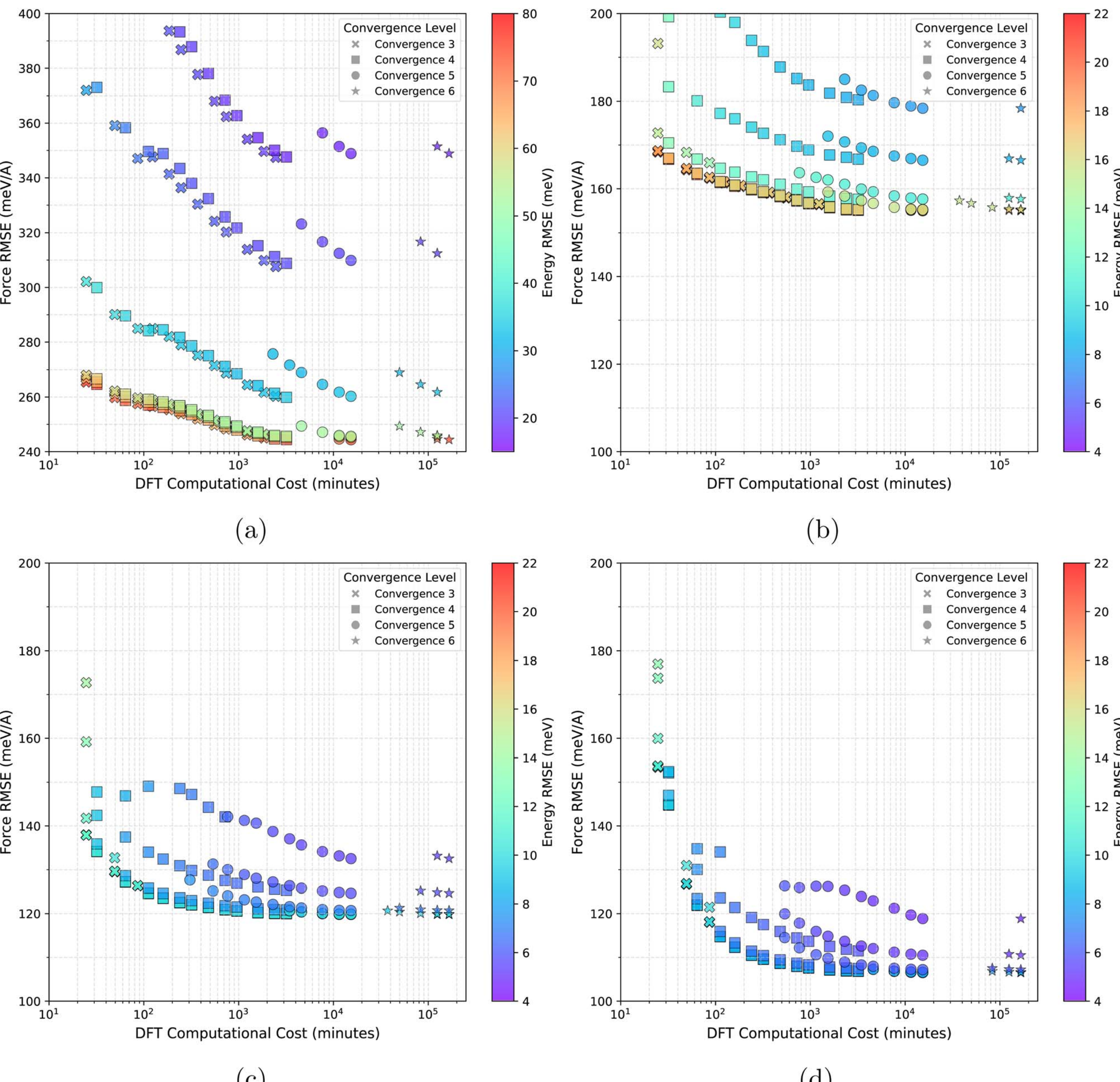

Fig. 8 Testing RMSEs (force *vs.* DFT time) for Pareto-optimal MLIP models of varying complexity ($2J_{\max} \in \{4, 6, 8, 10\}$); all models were tested on the level 6 convergence DFT testing set. Marker color indicates the energy testing RMSE. Marker shapes distinguish the DFT convergence levels used in training. Markers are organized into lines, where each line corresponds to a specific energy–force weight ($w_E/w_F \in \{5, 10, 12.25, 50, 150, 300\}$). Along each line, individual markers denote different training set sizes, ranging from 100 to 10 000 configurations. (a) $2J_{\max} = 4$, (b) $2J_{\max} = 6$, (c) $2J_{\max} = 8$, and (d) $2J_{\max} = 10$.

Further analysis of Fig. 7, 8 and S8 reveals that the optimal DFT level convergence depends strongly on both the MLIP complexity ($2J_{\max}$) and whether energy or force accuracy is prioritized. For complex MLIPs ($2J_{\max} = 10$), the limiting energy and force errors saturate around 4.5 meV per atom and 105 meV Å$^{-1}$ respectively. In this case, it is possible to approach both of these limits simultaneously through a proper choice of the energy/force weights. Approaching limiting force errors (*c.f.*, Fig. 8d) is possible even with lower convergence training sets, *e.g.*, the level 4 convergence, while approaching the ultimate energy error level requires convergence level 5 training sets (*c.f.*, Fig. 8d), which is consistent with the observation that the comparatively unbiased statistical properties of the forces make them more susceptible to being averaged out. Nonetheless, the effect is relatively modest in absolute terms, as even convergence level 4 training sets produce MLIPs whose errors are within 1.5 meV per atom and 5 meV Å$^{-1}$ of the saturation limit. Perhaps most surprising is the observation that very good MLIPs can be obtained very efficiently. For example, models with energy and force errors within 3 meV per atom and 20 meV

Å$^{-1}$, respectively, of the saturation limit can be obtained with only approximately 2 h of total computational time for DFT simulation using convergence level 4. In fact, errors show a very fast decrease in the first hour of DFT simulations, followed by a significant slowing down where further improvements come at a high computational cost. This shows that it is possible to obtain accurate models at an extremely low computational cost. Interestingly, the results also show that the optimal convergence level in fact depends on the total computational resources available for creating the DFT training set, as a small amount of computational resources (10–20 minutes) favors convergence level 3 training sets, intermediate budgets (20–1000 minutes) tend to favor convergence level 4 training sets, and large budgets (>3000 minutes) tend to favor convergence level 5 or rarely even level 6 training sets. Intuitively, this suggests that capturing the broad features of the energy landscape is better achieved with a larger number of low-convergence configurations in the training set than with a small number of high-convergence configurations in the training set, but that refining the detailed features of the landscape gradually calls for higher-convergence configurations in the training sets.

With low-complexity MLIPs ($2J_{\max} = 4$ and $2J_{\max} = 6$), limiting energy and force errors worsen significantly. Simultaneously approaching both limits using a simple energy–force weighting is no longer feasible due to a strong energy–force error trade-off. In contrast to high-complexity MLIPs, the decrease in errors is also significantly more gradual as more DFT simulations are added to the training set. For example, focusing on "intermediate" models that attempt to strike the balance between energy and force errors (*e.g.*, energy errors between 30 and 40 meV per atom and force errors between 260 and 300 meV Å$^{-1}$ for $2J_{\max} = 4$), one observes that increasing the computational resources for DFT simulation from $10^2$ to $10^3$ minutes leads to a substantial decrease in energy and force errors (in absolute terms) by about 7 meV per atom and 20 meV Å$^{-1}$, respectively, which is significantly larger than the corresponding decrease for $2J_{\max} = 10$. This is perhaps counterintuitive, as simpler models should require less data to constrain, and could therefore be expected to converge faster with increasing training set size. However, this intuition does not extend to misspecified models where no combination of free parameters can perfectly reproduce the training data. In this case, adding more configurations to an already large training set (*i.e.*, much larger than the number of adjustable parameters) can still significantly affect the MLIP's accuracy, even leading to worsening test errors in some cases,[18] a phenomenon that can also be observed here for energy errors at large force weights. Although the leverage sampling strategy should partially mitigate this trend by introducing the most influential points first, the slow convergence of energy and force errors with respect to the training set size is consistent with the behavior of highly misspecified models.

# 4 Discussion

The above results explore the trade-offs between computational cost and accuracy when fitting application-specific MLIPs. They provide guidance for selecting the level of DFT convergence and MLIP complexity based on the specific application of the MLIP and the acceptable computational cost to evaluate the MLIP for MD simulation. Our work shows that significant computational cost can be saved by considering multiple factors that can influence the quality of the MLIP. While the quantitative cost-benefit trade-offs we identified are specific to the electronically simple beryllium system, we expect the qualitative trends to be broadly applicable. For more complex metals prone to electronic convergence difficulties, the optimal balance between DFT convergence and model complexity might shift, but the existence of such an optimal, non-trivial balance remains a key principle.

First, we find that very tightly converging DFT calculations can be wasteful when coupled with MLIPs of limited expressivity. Table 4 and 5, along with Fig. 7, show that MLIPs trained on convergence level 5 training sets provide the same level of accuracy as those trained on convergence level 6 at a 10-fold reduction in cost. This is consistent with the rule of thumb that the accuracy of the model and of the data benefits from being roughly matched.

Second, further savings are possible by considering how DFT errors translate into MLIP errors, which is itself dependent on the expressivity of the model. Our Pareto analysis (Fig. 7) shows that if applications allow for more expressive MLIPs (*e.g.*, $2J_{\max} \geq 6$), training to medium convergence data (level 4 in our case) selected by leverage sampling can produce MLIPs nearly as accurate as those trained on convergence level 5 training sets. This approach can reduce the DFT computational cost by about 10 times compared to using the full convergence level 5 training set, and up to about 100 times compared to using the highest convergence level. Under optimal conditions, convergence with respect to the training set size can occur extremely quickly, using only a handful of GPU hours.

Third, if the application requires very computationally efficient MLIPs (*e.g.*, $2J_{\max} = 4$), these simpler MLIPs are less sensitive to DFT errors. Fig. 7d shows that in these cases, lower-convergence data can be used. This further reduces the computational cost of the DFT simulation while maintaining the same accuracy similar to higher-convergence training sets for the same simple MLIP. Using convergence level 3 saves about 50 times the computational cost for DFT simulation compared to convergence level 5, and over 500 times compared to convergence level 6 to characterize the whole training set. However, we also observed that very simple potentials can counterintuitively require a larger amount of data to achieve convergence with respect to the training set size due to misspecification effects.

Of course, when assembling datasets for applications that require accuracy at the expense of inference cost, or when curating databases of reference results, *e.g.*, to train universal models, erring on the side of caution and employing tight DFT convergence settings remain safe strategies. However, even in this setting, quickly generating lower-convergence data to fine-tune universal models might prove to be the superior option.

# 5 Conclusion

Developing cost-effective, accurate machine-learned interatomic potentials (MLIPs) remains essential, particularly for specific applications requiring computationally efficient MLIPs for long/large Molecular Dynamics (MD) simulation or in cases where multiple different materials must be considered on a limited computational budget. In order to delineate the importance of the different factors that contribute to the cost/accuracy tradeoff of MLIPs, we considered the role of DFT training set convergence, MLIP model complexity, training set sub-sampling, and relative energy/force reweighting. Our study demonstrates the benefits of leveraging lower-convergence DFT training sets, which are computationally much more efficient to obtain than the typical high-convergence DFT advocated in the literature. However, efficiently leveraging lower-convergence training sets is shown to require the consideration of the relative weighting of energies and forces in the overall loss function, as DFT errors in the forces converge faster than errors in the energy.[43] Coupled with training set sub-selection strategies based on leverage sampling, we mapped out the Pareto front of locally optimal MLIPs with respect to energy and force errors, and the computational cost to construct the DFT training set. The results confirm that the use of lower convergence DFT training sets can be beneficial when the convergence of the DFT training set can be roughly matched to the intrinsic levels of errors due to finite complexity of the MLIP. This is especially true for constrained computational budgets where lower convergence DFT training sets dominate the Pareto front. Through a careful optimization of the DFT convergence, energy/force weights, and training set size, we observed that MLIPs for unary beryllium that approach the intrinsic error saturation limit by a few meV per atom and meV $Å^{-1}$ respectively can be obtained with as little as 2 hours of aggregate computational resources for DFT simulation using medium convergence training sets, but that fully converging the MLIPs can require orders of magnitude more computational resources. Counterintuitively, we also observed that converging low-complexity MLIPs can be even more costly in terms of computational resources for DFT simulation, due to slow convergence with respect to the training set size, which we attribute to the effect of model misspecification. Our study suggests that extremely large efficiency gains can be achieved through a joint consideration of the multiple factors that control the cost/accuracy trade-off of MLIPs, highlighting the advantages of application-specific MLIPs.

# Author contributions

Ilgar Baghishov: methodology, software, data curation, writing – original draft. Jan Janssen: conceptualization, methodology, data curation, writing – review & editing, supervision, funding acquisition. Graeme Henkelman: methodology, writing – review & editing, supervision, funding acquisition. Danny Perez: conceptualization, methodology, writing – review & editing, supervision, funding acquisition.

# Conflicts of interest

There are no conflicts to declare.

# Data availability

The codes used in this work are available on GitHub and have been archived on Zenodo (https://doi.org/10.5281/zenodo.15814503). These scripts allow for the training of MLIPs in a full factorial design, systematically varying parameters such as subset size, energy/force weight, model complexity, and DFT convergence, followed by a Pareto analysis. The repository also contains the beryllium dataset generated with the maximum entropy method, which includes the DFT energies, forces, and stresses for all six convergence levels described in this paper.

Supplementary information is available. See DOI: https://doi.org/10.1039/d5dd00294j.

# Acknowledgements

I. B. acknowledges support from Machine Learning Fellowship program at Los Alamos National Laboratory and from the US Department of Energy through the Exascale Computing Project (17-SC-20-SC), a collaborative effort of the U.S. Department of Energy Office of Science and the National Nuclear Security Administration. D.P. gratefully acknowledges support from the U.S. Department of Energy, Office of Fusion Energy Sciences (OFES) under Field Work Proposal Number 20–023149. G.H. gratefully acknowledges support from the CCI grant CHE-2221062. J. J. gratefully acknowledges funding from the Deutsche Forschungsgemeinschaft (DFG) through the CRC1394 "Structural and Chemical Atomic Complexity – From Defect Phase Diagrams to Material Properties", project ID 409476157. I. B. thanks Sung Hoon Jung for sharing his expertise in leverage sampling. The authors thank Ivana Gonzales for the support in preparing DFT data sets at various convergence levels. Finally, the authors acknowledge the hospitality of the Institute of Pure and Applied Mathematics (IPAM) as part of the "New Mathematics for the Exascale: Applications to Materials Science" long program where this study was initially conceived. Los Alamos National Laboratory is operated by Triad National Security, LLC, for the National Nuclear Security Administration of U.S. Department of Energy (Contract No. 89233218CNA000001).